\documentclass[sigconf, acmReferenceFormat=false]{acmart}
\AtBeginDocument{%
  }

\copyrightyear{2025}

\acmConference[ ]{ }{ }{ }

\renewcommand\footnotetextcopyrightpermission[1]{}
\begin{document}

\title[Draw an Ugly Person]{“Draw an Ugly Person”: An Exploration of Generative AI’s Perceptions of Ugliness}

\author{Huisung Kwon}
\authornote{All authors contributed equally to this research}
\authornote{Department of Industrial Design}
\email{huisungk@kaist.ac.kr}
\affiliation{Korea Advanced Institute of Science and Technology \country{Daejeon, South Korea}}

\author{Garyoung Kim}
\authornotemark[1]
\authornote{Department of Culture Technology}
\email{grkim1507@kaist.ac.kr}
\affiliation{Korea Advanced Institute of Science and Technology \country{Daejeon, South Korea}}

\author{Seoju Yun}
\authornotemark[1]
\authornotemark[2]
\email{zupam@kaist.ac.kr}
\affiliation{Korea Advanced Institute of Science and Technology \country{Daejeon, South Korea}}

\author{Yu-Won Youn}
\authornotemark[1]
\authornotemark[3]
\email{yuwon@kaist.ac.kr}
\affiliation{Korea Advanced Institute of Science and Technology \country{Daejeon, South Korea}}


\begin{abstract}
Generative AI does not only replicate human creativity but also reproduces deep-seated cultural biases, making it crucial to critically examine how concepts like “\textit{ugliness}” are understood and expressed by these tools. This study investigates how four different generative AI models understand and express ugliness through text and image and explores the biases embedded within these representations. We extracted 13 adjectives associated with ugliness through iterative prompting of a large language model and generated 624 images across four AI models and three prompts. Demographic and socioeconomic attributes within the images were independently coded and thematically analyzed. Our findings show that AI models disproportionately associate ugliness with old white male figures, reflecting entrenched social biases as well as paradoxical biases, where efforts to avoid stereotypical depictions of marginalized groups inadvertently result in the disproportionate projection of negative attributes onto majority groups. Qualitative analysis further reveals that, despite supposed attempts to frame ugliness within social contexts, conventional physical markers such as asymmetry and aging persist as central visual motifs. These findings demonstrate that despite attempts to create more equal representations, generative AI continues to perpetuate inherited and paradoxical biases, underscoring the critical work being done to create ethical AI training paradigms and advance methodologies for more inclusive AI development.
\end{abstract}


\settopmatter{printacmref=false}
\setcopyright{none}
\maketitle
\section{Introduction}
\subsection{Background}
Diverse forms of content, including images, text, and video can now easily be generated by AI with the recent rise of generative AI models. These models demonstrate remarkable expressive capabilities learned from vast datasets; yet they inevitably recombine and replicate the language, visual codes, and value systems embedded within human societies. AI-generated outputs often reflect and even amplify existing social biases, stereotypes, and patterns of discrimination, thereby mirroring (and at times even exacerbating) broader societal structures \cite{inbtro1, Baxter_2023, intro_2024, zhou2023bias}. 

This dynamic becomes particularly complex when making aesthetic judgements. Aesthetic evaluation is largely subjective and context-dependent, shaped by cultural, emotional, and historical factors \cite{yang2025perspectives}. Perceptions of beauty and ugliness vary dramatically across individuals and societies, rendering them resistant to standardized or objective measurement. Nevertheless, generative AI systems attempt to encode and generalize such subjective value judgments through their training on large-scale datasets. In doing so, these systems run the risk of reinforcing dominant aesthetic norms and reproducing social biases without critical mediation \cite{inbtro1, intro_2024}. 

While previous studies have attempted to assess the aesthetic dimensions of AI-generated content, they concentrated on positive aesthetic values such as beauty, appeal, and visual attractiveness \cite{goring2023analysis, stacy2025representation, yang2025racial}. Relatively little attention has been paid to aesthetic categories with negative connotations, particularly perceptions of “ugliness,” and how they are conceptualized and materialized by generative AI systems. Biases are not limited to the value assignment of positive traits. They are equally embedded in the stigmatization of perceived flaws, deviations, and abnormalities. Without critical examination of how AI systems internalize and reproduce negative aesthetic judgments, we risk overlooking the ways in which exclusionary social values are perpetuated and potentially amplified through new forms of technological mediation.

\subsection{Importance and Research Questions}

This study aims to critically examine how generative AI perceives, conceptualizes, and expresses the notion of \textbf{\textit{ugliness}}. Ugliness is not merely the absence of beauty, but constitutes a complex, culturally and historically contingent value shaped by emotional, social, and aesthetic forces. Investigating how AI systems materialize this multifaceted concept offers a powerful lens through which to interrogate how aesthetic norms, social biases, and mechanisms of exclusion are algorithmically internalized and reproduced. We pose the following research questions to unpack this notion: 

\begin{quote}
    \textbf{\textit{RQ1: How does generative AI interpret and express “ugliness” both visually and linguistically?}} 
\end{quote}

\begin{quote}
    \textbf{\textit{RQ2: What do generative AI models' depictions of "ugliness" reveal about their underlying biases?}} 
\end{quote}

By addressing these questions, this study aims to contribute to a more critical understanding of the social and ethical implications of generative AI. We highlight the need to move beyond the evaluation of positive aesthetic values and to pay more attention to the ways negative aesthetic judgments are constructed, internalized, and disseminated by AI models. Through this inquiry, we seek to uncover the often invisible systems of exclusion that operate within generative technologies and to foster discussions toward developing more reflexive, culturally aware, and socially responsible AI practices. 
\section{Methodology}

\subsection{Data Collection}
\subsubsection{Extraction of adjectives reflecting ugliness}

\begin{figure*}[t]
    \centering
    \includegraphics[width=\linewidth]{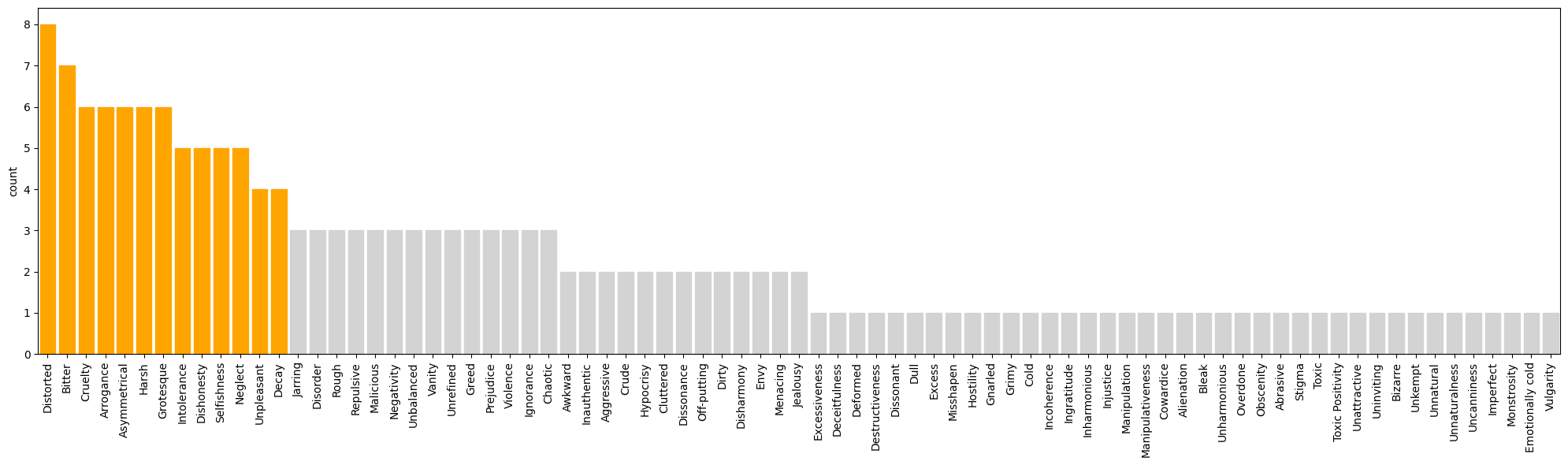}
    \caption{List of words associated with ugly as generated by ChatGPT}
    \label{tableadj}
\end{figure*}

The notion of ugliness is shaped by cultural, social, and emotional factors, making it difficult to define objectively. Therefore, we began our study with an exploration of what values or characteristics generative models associates with ugliness and generated images based on those adjectives. 

We prompted ChatGPT to generate 15 values or characteristics associated with the term \textit{ugly}. This process was repeated 12 times, yielding a total of 180 words. From here, 13 words that appeared most frequently (at least four times) were selected for the image generation process. Selected words include Arrogant, Bitter, Cruel, Dishonest, Distort, Harsh, Intolerance, Neglectful, Selfishness, Grotesque, Asymmetrical, Decaying, and Unpleasant. (Figure \ref{tableadj}, Appendix \ref{adjectives})

\subsubsection{Image Generation with Generative AI Models}
While ugliness is a multifaceted concept encompassing moral, emotional, physical, and environmental dimensions, exploring all these layers simultaneously risks blurring interpretations of the generated images. To mitigate this, our team decided to focus on representations of an “ugly person” as an initial exploration in this direction. We generated images via four commonly used generative AI models capable of image generation (ChatGPT, Grok, Midjourney, and Gemini) using the 13 words gathered in the earlier step to depict different aspects of \textit{\textbf{ugliness}}. A total of 624 images were generated in our experiment (13 adjectives x 3 prompts x 4 models x 4 images each). The generated images were generated April 7-10, 2025 and can be viewed through the link in the Appendix \ref{figma}. 

The three prompts used to generate images were designed to focus on different aspects of AI’s interpretation of \textit{\textbf{ugliness}}. We asked AI models to “Draw a(n) [ugly adjective\footnote{One of the 13 different words previously extracted in 2.1.1}] person” (\textit{Prompt A}) as a way to examine how each adjective is visually interpreted by AI models, then asked AI models to “Draw a(n) [antonyms of ugly adjective\footnote{The antonyms corresponding to the ugly adjectives are listed in Appendix A. These words were selected based on dictionary definitions and cross-validated by a bilingual author.}] person” (\textit{Prompt B}) as an attempt to critically explore AI’s interpretation of ugliness through comparison. We hoped to see what choices were being made differently for words carrying more positive or negative connotations. The final image generation prompt used was “Within the context of ugliness (however you define it), draw a(n) [ugly adjective] person” (\textit{Prompt C}). This prompt asks what the generative AI models consider \textit{\textbf{ugly}} in a more direct way. Since the first two prompts do not directly ask the model to draw an “ugly person,” but rather a person with adjectives associated with ugliness, this supplementary prompt was necessary to examine whether the generated images were indeed contextualized within the context of ugliness and not just the dictionary definition of the adjectives. We assumed that the differences observed between the two groups of images—those generated from Prompt A and those from Prompt C—would reveal the core of how generative AI depicts ugliness. Additionally, we requested the AI models to “explain why you drew the picture like that,” with the aim of making AI verbalize their underlying judgments or value systems when creating images for Prompt C. 

In order to maintain control and limit the effect of personalized results, temporary chats were used when possible and new chats were created for each image generation. 

\subsubsection{Data Constraints and Limitations}
Due to external factors such as the limitations of the AI models used, we faced some data constraints.  First, although “clear” was initially selected as the antonym of “distorted,” the AI models produced images resembling invisible figures (e.g., figures made of water) due to the multiple definitions of the word and the limited context we provided. This was determined as inappropriate for the subsequent analysis. The word “perfect” was retrospectively added as an antonym of “distorted” to generate images to mitigate this issue. Second, linguistic explanations for Prompt C were collected from only three generative AI models (ChatGPT, Grok, and Gemini) due to Midjourney's inability to generate text.

\subsection{Data Analysis} 
\subsubsection{Attribute coding and quantitative analysis of generated images}
To examine the biases within the generated images, we conducted attribute coding for each image across perceived visual, cultural, and socioeconomic layers through five attributes: gender, age, ethnicity, hygiene, and clothing. Due to time constraints, encoding was done by the four authors who independently coded the attributes of each image, then cross-validated to ensure objectivity to the best of our ability. We acknowledge that all coding decisions were based on the authors’ interpretations of a given image and the perceived attributes may not be the most accurate reflection of the individual's characteristics within the image. 

Perceived gender refers to the gender of the figure as interpreted by the researchers, categorized as perceived male, perceived female, or undefined if it could not be clearly identified. Perceived age was tagged based on visual indicators such as wrinkles and gray hair. Individuals were categorized as child/adolescent, young adult, middle-aged, senior, or undefined if their age could not be easily determined. Perceived ethnicity was categorized according to the classification standards established by the U.S. Census Bureau and the Office of Management and Budget (OMB) Statistical Directive No. 15. Categories included Multiracial/Mixed, Latin American, Asian, African, Indigenous, European, and Middle Eastern/North African. Individuals whose ethnicity could not be easily determined were categorized as “Uncertain.” Additionally, applying the concept of “People of Color” (PoC), which is widely used in social science research to encompass nonwhite populations \cite{starr2024people}, we further classified perceived ethnicities into White, PoC, and Uncertain. Hygiene was classified into clean, neutral, and unclean categories based on visual indicators such as holes in clothing or trash in the environment. Finally, clothing style was categorized into formal,semi-formal, casual, sportswear, nonstandard, and uncertain. The latter two attributes were tagged to provide insight into socio-economic aspects of the images as objectively determining one’s wealth or class is difficult from a single image.

\subsubsection{Qualitative analysis of generated textual data}
The goal of our analysis was to examine the underlying judgments and value systems regarding \textbf{\textit{ugliness}} and to explore the attitudes embedded in the LLMs. Therefore, we also thematically coded the text data generated in Prompt C. We manually conducted the initial coding process after reading over all of the text and identifying recorruing themes, focusing on expressions that reflected social norms, cultural expectations, and negative value judgments. Based on the initial codes, similar or related codes were grouped to generate preliminary themes. Finally, the themes were reviewed, merged, and refined through the iterative refinement process, resulting in the two final themes.

\section{Analysis and Results}
\subsection{Results of Quantitative Analysis}
\subsubsection{Perceived Demographic Attributes}

\begin{figure*}
    \centering
    \includegraphics[width=0.8\linewidth]{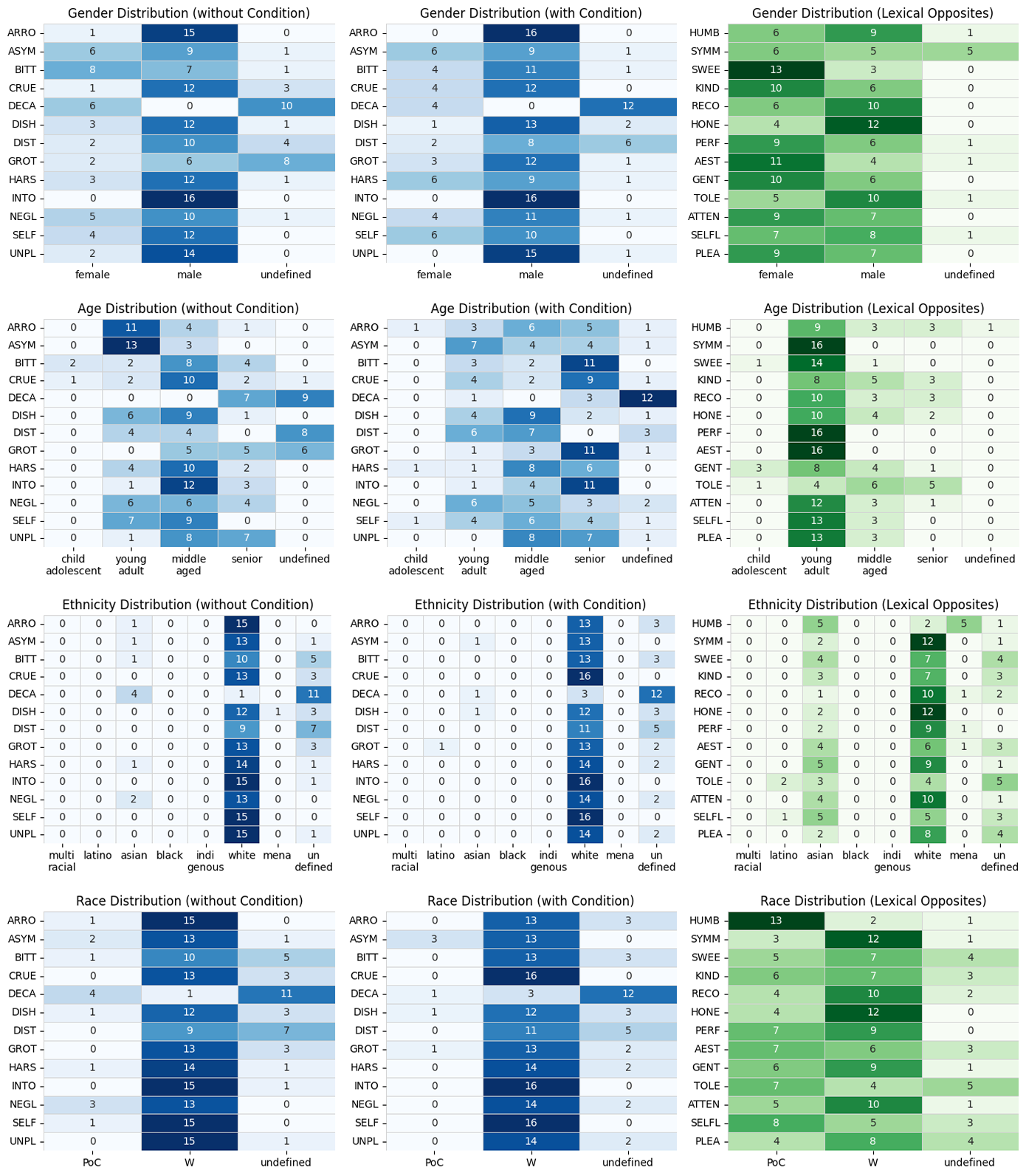}
    \caption{Visualized Demographic Attribute Distributions for "Ugly Person" Image Generations}
    \label{results1}
\end{figure*}

\paragraph{Gender}
Prompt A (without context) resulted in an over-representation of men for most adjectives with more than 14 out of 16 images being classified as male for three adjectives (arrogant, intolerant, unpleasant). Some adjectives saw a more even split (asymmetrical, bitter), but men were still the dominant gender present with only two exceptions of the decaying and grotesque. Decaying is a clear outlier due to most images being undefinable while the people with identifiable gender were perceived as female. 
This tendency to over represent people perceived as male was also found in the results of Prompt C (with the context of ugliness). The most notable difference in gender attribution with the addition of the context of ugliness was in the adjectives bitter and grotesque as the former’s gender split shifted towards a more even divide (8 female, 7 male vs 4 female, 11 male) and the latter resulted in more people with undefined gender (2 female, 6 male, 8 undefined vs 3 female, 12 male, 1 undefined). 
Conversely, images generated with Prompt B (antonyms for ugliness) resulted in a clear difference in gender distribution. The antonyms sweet, kind, perfect, aesthetic, gentle, attentive, and pleasant resulted in more individuals perceived as female being drawn by the generative AI models while symmetrical and selfless resulted in an almost even distribution. 

Though there are adjective specific nuances, the results suggest that there is an overall biased tendency to generate individuals perceived as male when expressing the idea of \textbf{\textit{ugliness}} in an image of a person.

\paragraph{Age}

In Prompt A, the most common age range for half of the adjectives (bitter, cruel, dishonest, harsh, intolerant, selfish, unpleasant) was middle aged while drawing arrogant and asymmetrical people resulted in mostly young adults with 11 and 13 images respectively. The other adjectives resulted in more instances of individuals whose age could not be easily identified. 

When the context of ugliness was added in Prompt C, there was a change in the most commonly seen age group for most adjectives (all except decaying, distressed, and neglectful). For each, the shift tended towards higher frequencies of older people than instances of the same adjective in Prompt A. For example, arrogant had 11 young adults in Prompt A results, but that number dropped to 3 young adults in Prompt C while the number of seniors rose from 1 to 5. A similar trend can be seen throughout the adjectives for ugliness revealing that the context of ugliness leads to generations of older people.

Prompt B resulted in almost all generated images being young adults. The only adjective that did not have young adult as the most common age range was tolerant. These results show that age is an indicator of ugliness to generative AI with the age of visualized individuals changing for the older as more layers of ugliness are added and youth being a trait indirectly associated with positive connotations. 

\paragraph{Race and Ethnicity}

When looking at the coded results for race and ethnicity for Prompt A, only 14 total POC are present in the sample while there were at least 10 white people per adjective for 12 out of 13 adjectives. Taking a look at the perceived ethnicity of these images reveals that the majority of the POC were Asian. Prompt C did not generate significantly different results compared to Prompt A, though a few adjectives saw a slight increase in the number of white individuals. For three of those adjectives (cruel, intolerant, selfish), every image generated was of a white person. 

In contrast, looking at the results of Prompt B revealed a significant difference. White people were still the majority for half of the adjectives (symmetrical, recovering, honest, perfect, gentle, attentive, pleasant) in this prompt, but other adjectives saw less than half the results return white people. The ethnicity distribution of Prompt B’s results are also more diverse with instances of indigenous and multiracial individuals present. Though more POC were present, the most represented ethnicity within POC was consistent with the other prompts’ results with Asian being the most common ethnicity within POC generated.

These results show generative AI has a strong tendency to use white people as the default when visualizing people, but this tendency is even stronger when visualizing concepts of ugliness while racial diversity is disproportionately introduced for the positive antonyms (Table \ref{results1}).

\subsubsection{Perceived Socioeconomic and Lifestyle Attributes}

\begin{figure*}
    \centering
    \includegraphics[width=0.8\linewidth]{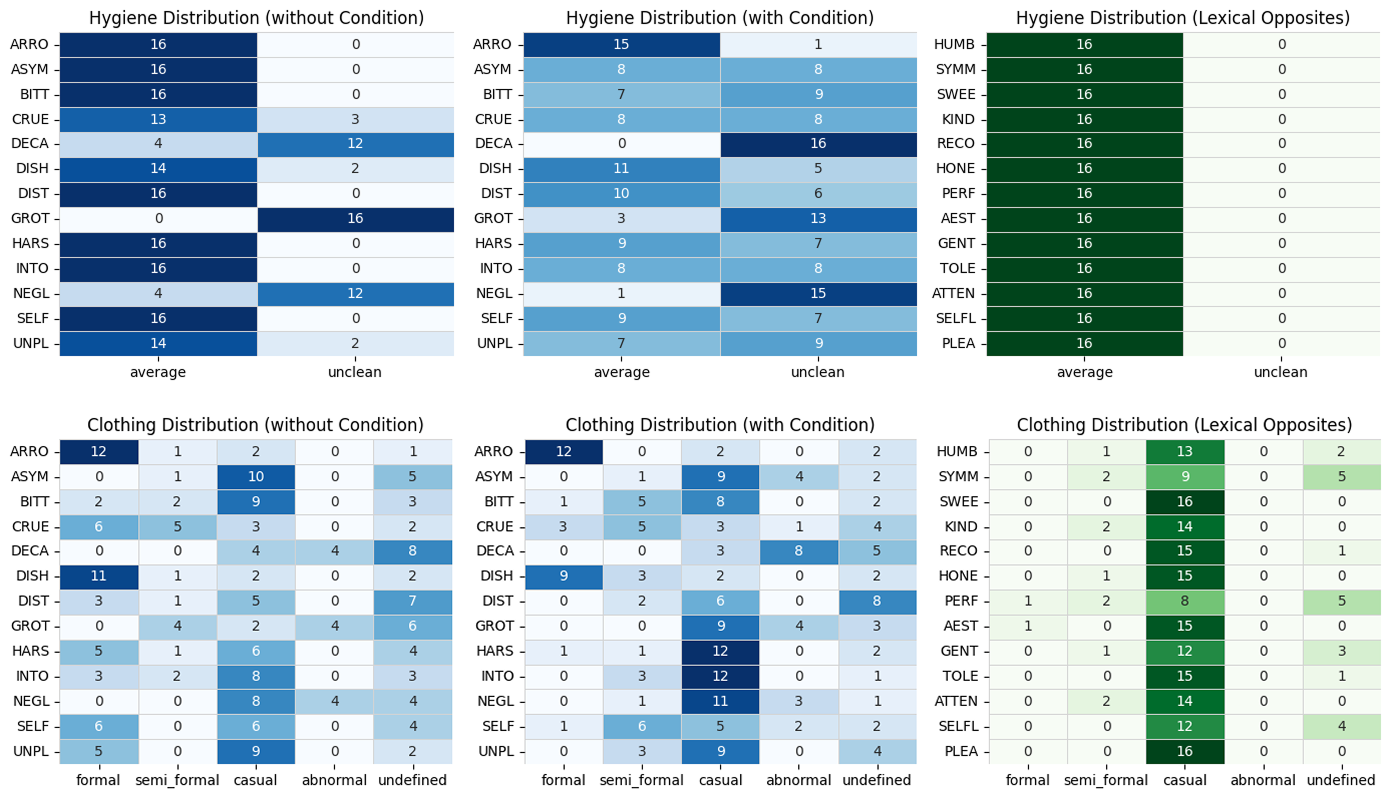}
    \caption{Visualized Socioeconomic and Lifestyle Attribute Distributions for "Ugly Person" Image Generations}
    \label{results2}
\end{figure*}

\paragraph{Hygiene}
When looking at the hygiene levels of the people generated in Prompt A, most adjectives resulted in normal levels of hygiene. The exceptions were decaying, grotesque, and neglectful, which resulted in images of mostly unclean people. Prompt C’s results were quite distinct from that of Prompt A. Adjectives that were mostly average levels of hygiene were split almost evenly across average and unclean hygiene levels. Adding the context of ugliness led to a decrease in hygiene levels. 
Prompt B reveals a similar trend in the opposite direction. The antonyms resulted in no adjectives drawing unclean individuals. Turning away from ugliness led to an increase in cleanliness. 

The hygiene levels observed across the prompts show that generative AI does have a tendency to see ugliness as dirtier than its counterparts. However, since the shift is only towards a moderate split between average and unclean, it does not appear that low hygiene levels are an essential part of ugliness for generative AI.

\paragraph{Clothing}
Prompt A and C revealed similar results in terms of clothing generated within images of ugliness. Formal clothing is seen frequently in images of arrogant and dishonest people while most adjectives result in images containing people in casual clothing. In contrast, the antonyms in Prompt B resulted in all adjectives drawing images with people in casual clothes.

These results imply that generative AI does not place particular emphasis on clothing when visualizing ugliness. As the type of clothing most commonly seen is maintained across prompts, it can be inferred that the expression of clothing can be interpreted as reflecting the nature of the adjective used rather than the value judgment of ugliness (Table \ref{results2}).

\subsection{Findings of Qualitative Analysis}
\subsubsection{Expression of Visual Ugliness in Words}
We noticed two main trends in LLM’s textual explanations of ugliness. First, many responses included contextual elements such as lighting, background, posture, and attire in representations of ugliness despite the prompt asking for a drawing of an ugly person. Examples included dim or moody lighting, blurred or undefined spaces, slouched body posture, and unkempt appearance. These elements appeared regardless of whether physical deformities were also present.

\begin{quote}
    \textit{“The palette is muted and cold, stripped of warmth, suggesting emotional distance. The lighting is stark and directional—like an interrogation lamp—emphasizing her severe presence and every line of experience, regret, or resentment.”} (ChatGPT) 
\end{quote}

Second, physical features traditionally associated with unattractiveness such as wrinkles, asymmetry, sagging facial lines, and rough skin texture were often framed as markers of life’s weight and trajectory and not simply flaws in appearance. The continued addition of context to the emphasis on these visual characteristics suggests an attempt to portray physical irregularities not as superficial flaws, but as reflections of accumulated experiences and inner turmoil. This can be seen in prompts focused on emotional or behavioral ugliness in particular.

\begin{quote}
    \textit{“For this image, I defined "ugliness" through exaggerated aging and emotional distress, distorting the person’s appearance to reflect inner turmoil.”} (Grok) 
\end{quote}

\subsubsection{Avoidance of accountability}

When asked to explain why they drew an ugly person the way they did, the various generative AI models exhibited a tendency to make attempts to avoid accountability, employing various tactics to do so. 

One tactic noticed was a deliberate distancing of itself from the decision-making behind image generation by shifting blame to the data or a separate image generation model. Statements such as “This is based on my understanding of the prompt and training data” appeared across multiple responses and models while Gemini explicitly stated that “I cannot know for sure why the image was generated the way it was, but I can offer some possible explanations” in another instance.  Instances of chatbots (Gemini and ChatGPT) claiming that they could not know the reasoning as image generation occurs in a separate model (Imagen and Dall-E) were recorded as well.

In other cases, the AI models redirected the judgment of ugliness back to the user, prompting reflection rather than asserting a fixed standard:

\begin{quote}
    \textit{“The image invites viewers to confront their own biases and preconceived notions about what is considered "ugly" and why.”} (Grok) 
\end{quote}

This tactic of passing the locus of interpretation onto the viewer was observed consistently across different prompt types and models, suggesting a deliberate strategy to avoid direct value judgments while still guiding users toward self-examination. The generative AI models do not claim to show something ugly in these instances, but instead tell the viewer that the ugliness is something they find within an image that isn’t necessarily one thing or another.  According to the models that provide depictions of ugliness, they are not making any claims of what is ugly or beautiful in their generation of such images.

\section{Discussion}
\subsection{The Possibility of Reverse Discrimination}
In this study, we observed a tendency for ugliness to be primarily represented visually by white men. This could potentially be attributed, first, to the demographic composition of the training dataset. Large-scale datasets tend to over represent image and text data related to white men \cite{aldahoul2024}, which may lead to frequent depictions of white men during the generation process. Additionally, the algorithm may have undergone post-training alignment \cite{kumar2025llmposttrainingdeepdive} to ensure a positive balance for minority groups when assigning favorable traits (like beauty). As a result, there's a possibility that the negative trait of "ugliness" was disproportionately projected onto majority groups (especially white men) in an effort to avoid reproduction of societal discrimination. The model may have intentionally avoided negative portrayals of certain groups (minorities) while excessively attributing negative traits to other groups (majorities), resulting in the biases observed in this study.

\subsection{Inheritance of Human Perceptions of Ugliness}
The way AI models depict values largely reflects the aesthetic and ethical judgments embedded in human society \cite{dis2}. This can be seen in our findings when comparing images created within the context of "ugliness" with those that were not where being older was correlated with being uglier. We must carefully consider whether we can simply label these characteristics of AI as "wrong." AI reproduces ugliness based on the given data and social contexts, which reflect the values and biases already present in human society. 

At the same time, the fact that AI consistently portrays natural physical changes, such as aging, in a negative aesthetic light risks reinforcing discriminatory perspectives in society and spreading negative perceptions about particular groups (e.g., the elderly, individuals with physical disabilities). Critically analyzing the depiction of ugliness expressed by generative AI goes beyond just pointing out technical errors. It is also about understanding how technology can amplify societal biases if left unchecked and working toward developing more responsible value judgment frameworks in the design and creation of AI systems.

\subsection{Contradictions of Generative AI}
When generating images representing "ugliness," AI systems primarily relied on conventional physical attributes from persistent appearance-based social stereotypes such as signs of aging, facial asymmetry, and rough skin texture. However, when prompted further, the AI explained that it sought to convey feelings of discomfort or social isolation through elements like lighting, facial expression, posture, and background with no mentions of the conventional physical attributes of ugliness utilized. This construction of an emotional narrative instead of a description of the main visual features of the person (the main point of the original prompt) appeared to supplement and retrospectively justify the visual depictions generated. 

A similar contradiction was observed in the text itself as well. Rather than employing direct terminology such as "ugly," the AI consistently relied on ambiguous expressions like "uncomfortable" or "emotionally distant." While this approach may mitigate issues of political sensitivity, it also blurs the transparency of the underlying aesthetic criteria and avoids accountability.
In this way, the AI’s approach both reflects sensitivity to ethical concerns and, at the same time, a desire to not engage directly with these issues.

\section{Limitations and Future Works}

This study was primarily conducted using English-language prompts and AI models trained on data developed in Western contexts. While this ensured linguistic consistency, it also introduced cultural bias, as the aesthetic representations of "ugliness" may reflect Western-centric values. To examine this, we conducted follow-up trials using prompts translated into Korean and Spanish. However, these versions produced images that were largely consistent with the English outputs, suggesting a strong underlying influence of model training data. Nonetheless, future research should further explore how culturally grounded prompts and AI models developed and trained outside of this context may generate different aesthetic outcomes.

The scope of our analysis was deliberately focused on human representations in order to maintain depth and coherence across cases. As a result, other potentially rich domains such as depictions of ugliness in environments, objects, or abstract concepts were excluded. Including these in future studies could offer a broader understanding of how AI interprets and visualizes negative attributes beyond the human form and outside of explicit socioeconomic structures.

Finally, our qualitative analysis involved inherently subjective interpretation. Despite having multiple coders involved in reviewing the dataset, traits such as "cleanliness" or "ethnicity" are context-dependent and may carry different cultural or personal connotations. To strengthen generalizability and reduce interpretive bias, future work could incorporate quantitative methods such as surveys or crowd-sourced evaluations across culturally diverse populations.

\section{Conclusion}
Through a quantitative and qualitative exploration of generative AI’s “ugliness,” we discovered assumptions and contradictions built into current generative AI models on the market. Images of “ugly” people were biased, but generative AI models sought to avoid accountability when prompted to explain the reasoning for its visual choices. Visual representations revealed the assumptions generative AI feels comfortable making while textual explanations revealed an attitude that runs contradictory to the assumptions baked into its own images. The realm of “ugliness” is an underexplored topic that can reveal much about the limitations of generative AI’s sensitivity protocols as well as the ways bias can manifest differently across various axes.

\bibliographystyle{ACM-Reference-Format}
\bibliography{references}

\appendix
\section{Appendix}

\subsection{Adjectives associated with ugliness and their antonyms} 
\label{adjectives}

\begin{enumerate}
    \item Arrogant — Humble
    \item Bitter — Sweet
    \item Cruel — Kind
    \item Dishonest — Honest
    \item Distort — Perfect
    \item Harsh — Gentle
    \item Intolerance — Tolerant
    \item Neglectful — Attentive
    \item Selfish — Selfless
    \item Grotesque — Aesthetic
    \item Asymmetrical — Symmetrical
    \item Decaying — Recovering
    \item Unpleasant — Pleasant
\end{enumerate}

\subsection{Link to Generated Images} 
\label{figma}
All generated images can be viewed at the FigJam link provided below. 

https://www.figma.com/board/WLatTW7GYzBVveyKbQWyZw/Draw-an-ugly-person?node-id=3-302\&t=uvqPUomGOBqNg5Nk-1

\end{document}